\documentclass{aip-cp}

\usepackage[numbers]{natbib}
\usepackage{rotating}
\usepackage{graphicx} 
\usepackage[dvipdfmx]{}
\usepackage{siunitx}

\begin{document}

\setlength\abovecaptionskip{0pt}

\title{Development of a low-alpha-emitting $\muup$-PIC for NEWAGE direction-sensitive dark-matter search}

\author[aff1]{Takashi Hashimoto\corref{cor1}}
\author[aff1]{Kentaro Miuchi}
\author[aff1]{Kiseki Nakamura}
\author[aff1]{Ryota Yakabe}
\author[aff1]{Tomonori Ikeda}
\author[aff1]{Ryosuke Taishaku}
\author[aff1]{Miki Nakazawa}
\author[aff1]{Hirohisa Ishiura}
\author[aff1]{Atsuhiko Ochi}
\author[aff1,aff2]{Yasuo Takeuchi}


\affil[aff1]{Department of Physics, Kobe University, Kobe, Hyogo 657-8501, Japan}
\affil[aff2]{Kavli Institute for the Physics and Mathematics of the Universe (WPI), The University of Tokyo Institutes for Advanced Study, University of Tokyo, Kashiwa, Chiba 277-8583, Japan}
\corresp[cor1]{takashi-hashimoto@stu.kobe-u.ac.jp}

\maketitle

\begin{abstract}
NEWAGE is a direction-sensitive dark-matter-search experiment that uses a micro-patterned gaseous detector, or $\muup$-PIC, as the readout.
The main background sources are $\alphaup$-rays from radioactive contaminants in the $\muup$-PIC. 
We have therefore developed a low-alpha-emitting $\muup$-PICs and measured its performances.
We measured the surface $\alphaup$-ray emission rate of the $\muup$-PIC in the Kamioka mine using a surface $\alphaup$-ray counter based on a micro TPC. 
 \end{abstract}

\section{INTRODUCTION}
A large fraction (\textasciitilde26\%) of the universe is in the form of nonbaryonic cold dark matter\cite{Planck}.
A weakly interacting massive particle (WIMP) is a well-motivated dark-matter candidate.
Numerous WIMP-search experiments have attempted to detect direct interactions between WIMPs from the galactic halo and target nuclei in a detector.
Despite numerous experimental efforts, no experimental evidence for the detection of WIMPs is found, except for an indication of an annual-modulation signal reported by the DAMA/LIBRA experiment\cite{DAMA_2012}, claimed as a positive signature of WIMPs.
However, several other experiments have obtained lower cross-section limits\cite{XENON}\cite{LUX}\cite{PANDA}.
Because the amplitude of the annual-modulation signal is only a few percent, positive signatures of WIMPs are very difficult to detect.
We require more robust evidence before definitively concluding that WIMPs exist. 
Direction-sensitive methods have been suggested to provide more convincing signatures for WIMPs\cite{DM_direction_strong_evidence}.
Because the Cygnus constellation is seen in front of the Solar system in its motion through the Galaxy, any galactic-halo WIMPs would appear to originate from the Cygnus direction as a ``WIMP-wind".
An asymmetric distribution of incoming WIMPs would provide strong evidence of WIMP detection. 
Direction-sensitive WIMP-search experiments measure both the energy and track of a recoil nucleus.
Among several methods being developed intensively around the world\cite{NIT_NIM2007}\cite{CNT_DM}, a gaseous time projection chamber using a low-pressure gas is considered most promising and has been studied for decades\cite{DRIFT_APP2012}\cite{MIMAC_2011}\cite{DMTPC_PLB2011}.

\section{NEWAGE}
NEWAGE is a direction-sensitive dark-matter-search experiment performed using a gaseous micro-time-projection chamber ($\muup$-TPC).
The $\muup$-TPC comprises  a two-dimensional, fine-pitch, imaging device called micro-pixel chamber ($\muup$-PIC) \cite{uPIC}, a gaseous electron multiplier (GEM)\cite{GEM}, and a detection volume (30 $\times$ 30 $\times$ 41 cm$^{3}$) filled with  CF$_4$ gas at 0.1 atm.
The area of the $\muup$-PIC is $30.72\times30.72\,\si{cm^{2}}$, corresponding to $768 \times 768$ pixels, with a pitch of $400\,\muup$m connected by 768 anode and 768 cathode strips.
The NEWAGE detector has been supplemented with a gas-circulation system using cooled charcoal to reduce radon contamination.
We used NEWAGE to perform a WIMP search in the Kamioka underground laboratory. 
The details, performance, and direction-sensitive limits of the NEWAGE detector are reported elsewhere\cite{NEWAGE_2015}.
We obtained a direction-sensitive SD cross-section limit of 557\,\si{pb} (90$\%$ confidence limit) for a WIMP mass of 200\,\si{GeV/c^{2}}\cite{NEWAGE_2015}.
Next, we plan to search the DAMA region using this direction-sensitive method, therefore, we need to improve the detection sensitivity from the current level.

%
%
\subsection{Background study}
To improve the sensitivity of our system, we need to understand and reduce the background sources. 
Accordingly, this background study aims to understand the measured energy spectrum. 
Radioactive contaminants within the detector components are well-known background sources in rare-event measurements.
In particular, many materials contain the natural radioactive isotopes $^{238}$U and $^{232}$Th.
These isotopes emit several $\alphaup$-rays, $\betaup$-rays and $\gammaup$-rays in their series decay.
\par
Using a high-purity germanium (HPGe) detector, we measured the contamination due to U/Th-chain isotopes in the materials employed in our detector components and there is some possibility that the decay rates of before (``upper stream”) and after (``middle stream") the ${}^{226}\rm{Ra}$ are inconsistent.
The half-life of ${}^{226}\rm{Ra}$ is much longer than that of the other isotopes in the U-chain.
What we term the ``$^{238}\rm{U}$ upper stream" comes from 93\,\si{keV} derived from ${}^{234}\rm{Th}$, whereas the ``$^{238}\rm{U}$ middle stream" originates from a peak at 609\,\si{keV} derived from ${}^{214}\rm{Bi}$.
We checked the decay rates for the upper and middle stream of U-chain and measured four different types of
samples: polyimide (PI) insulators, glass-cloth, plating solution and GEM.
The $\muup$-PIC contain three layers: a layer of polyimide (PI100 $\muup$m), PI800 $\muup$m and another PI100 $\muup$m.
We dissolved the PI out of a 100 $\muup$m insulator using NaOH and KOH and measured the activity of the remaining glass cloth. 
The measurement results are listed in TABLE \ref{tab:measurment_result}.
The results for the GEM and plating solution (CuSO$_{4}$) are below the measurement limit.
Only the PI100 $\muup$m and glass cloth contained in the PI100 $\muup$m layer have finite values.
We verified the radioactive equilibrium of the U series for these samples.
We converted the units from [$10^{-6}$ g/g] (TABLE \ref{tab:measurment_result}) into $[\si{\micro\becquerel/cm^2}]$ (TABLE \ref{tab:measurment_result2}).
 The results in TABLE \ref{tab:measurment_result2} show that the $^{238}$U and $^{232}$Th amounts in the PI100\,\si{\micro \metre} insulator can almost be explained by those in the glass cloth.

\begin{table}[ht]
\centering
\caption{Contamination by U/Th-chain isotopes in the materials of the $\muup$-PIC detector components, as measured using the HPGe detector. The uncertainties listed are statistical errors.}
\scalebox{0.9}{
\begin{tabular}{l|c|c|c}
\hline
Sample & $^{238}\rm{U}$ upper stream [$10^{-6}$ g/g] &$^{238}\rm{U}$ middle stream [$10^{-6}$ g/g] & $^{232}\rm{Th}$ [$10^{-6}$ g/g]\\ \hline \hline
PI100 $\muup$m insulator& $0.38\pm0.01$ &$0.39\pm0.01$ &  $1.81\pm0.04$\\
glass cloth &$0.91\pm0.02$ & $0.84\pm0.03$ & $3.48\pm0.12$\\ 
plating solution(CuSO$_{4}$) & $<0.13$ & $<0.01$ & $<0.06$\\ \hline
GEM & $<0.17$ & $<0.02$ &  $<0.12$\\
\hline
\end{tabular}
}
\label{tab:measurment_result}
\end{table}

\begin{table}[h]
\centering
\caption{Contamination by U/Th-chain isotopes in a sample PI100 $\muup$m insulator sheet. The uncertainties listed are statistical errors.}
\scalebox{0.9}{
\begin{tabular}{l|c|c|c}
\hline
Sample & $^{238}\rm{U}$middle stream $[\si{\micro\becquerel/cm^2}]$&$^{238}\rm{U}$ upper stream $[\si{\micro\becquerel/cm^2}]$&$^{232}\rm{Th}$ $[\si{\micro\becquerel/cm^2}]$\\ \hline \hline
PI100 $\muup$m insulator& $66.4\pm2.5$ & $68.5\pm1.5$ &  $102.1\pm2.3$\\
glass cloth & $71.3\pm1.4$&$64.5\pm0.8$ &$86.8\pm1.1$\\ 
\hline
\end{tabular}
}
\label{tab:measurment_result2}
\end{table}
\par
\begin{figure}[ht]
	 \centerline{\includegraphics[width=80mm,bb=0 0 842 595]{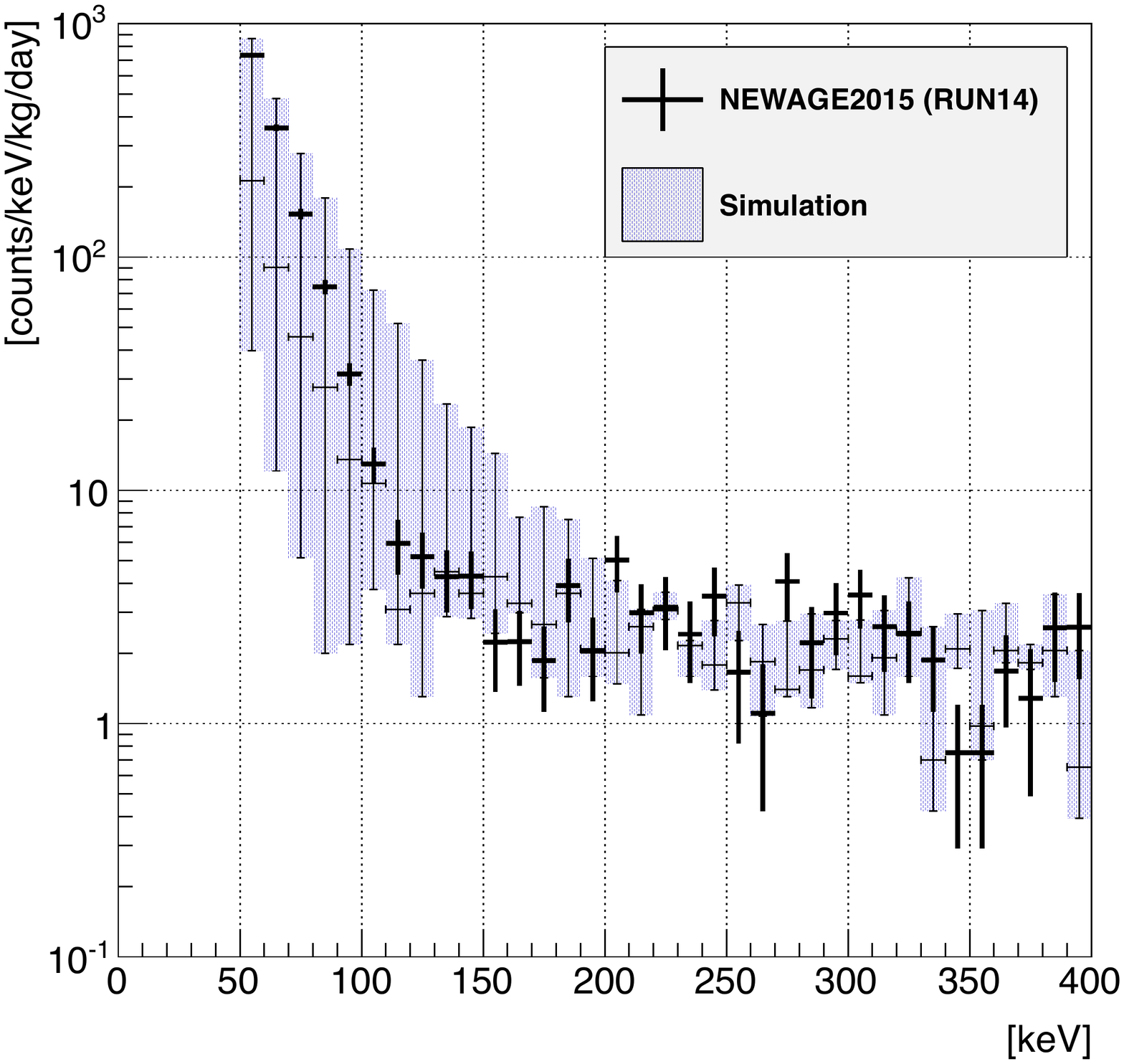}}
  \caption{Simulation results for the energy contributions of $\alphaup$-rays from $^{238}$U and $^{232}$Th in the reinforcing glass cloth inside the PI100 $\muup$m insulator. Black crosses are measurement results from NEWAGE2015, and the blue histograms are the simulation results. The band of the simulation histogram is dominated by the uncertainly in the glass-cloth thickness and the GEM gas gain.We compared the measured spectrum of NEWAGE2015 without the roundness-cut and simulation because the events from the substrate are basically cut by the effort of Z-fiducilization of roundness-cut.}
  \label{fig:sim_result}
\end{figure}

We simulated the energy contributions of $\alphaup$-rays from $^{238}$U and $^{232}$Th in the glass cloth inside a sample of PI100 $\muup$m insulator using the Geant4\cite{Geant4} using the measurement results from the $^{238}\rm{U}$ middle stream and $^{232}\rm{Th}$.
We compare the results of the NEWAGE2015 measurements and Geant4 simulation in Figure.\ref{fig:sim_result}.
We compared the measured spectrum of NEWAGE2015 without the roundness-cut and simulation because the events from the substrate is partially cut by the effort of Z-fiducilization of roundness-cut.
We consider uncertainly in the GEM gas gain and glass-cloth thickness to be systematic errors in the simulation results.
In the 50-150\,\si{\kilo\electronvolt} region, almost all events are not through the GEM hole. 
In this energy region, systematic errors in the uncertainly of the GEM gas gain are dominant.
In the 150-400\,\si{\kilo\electronvolt} region, almost all events are through the GEM hole. 
The dominant systematic errors here is uncertainly in the glass-cloth thickness.
The energy spectrum measured by NEWAGE2015 is thus well-explained by $\alphaup$-rays from $^{238}$U and $^{232}$Th the glass cloth inside the PI100 $\muup$m insulator.
Thus, we find the glass cloth used to reinforce the PI100 $\muup$m insulator as dominant background source.

\subsection{Development of a "Low $\alphaup$ $\muup$-PIC"}
The NEWAGE 2015 data showed that the dominant background source originates from the glass cloth in the PI 100$\muup$m insulator of the $\muup$-PIC.
We thus need to replace this insulator with a low-background material.
We selected a new material comprising of a 5-$\muup$m PI layer without reinforcing glass cloth and a 75-$\muup$m epoxy layer of glue.
The $^{238}$U and $^{232}$Th contamination measurement results in the new material (TABLE\ref{tab:LA_measurment_result}) show that it contains 100 times lower contamination than the current material. 
Accordingly, we decided to develop a new, low-$\alphaup$ $\muup$-PIC, by replacing the top layer of PI with the new material.

\begin{table}[ht]
\centering
\caption{$^{238}$U and $^{232}$Th measurement results using the HPGe detector. The uncertainties listed are statistical errors.}
\scalebox{0.9}{
\begin{tabular}{l|c|c|c|c}
\hline
Sample & $^{238}\rm{U}$upper stream [$10^{-6}$ g/g]&$^{238}\rm{U}$middle stream [$10^{-6}$ g/g] &$^{232}\rm{Th}$ [$10^{-6}$ g/g]& Note\\ \hline \hline
PI100 $\muup$m insulator& $0.38 \pm 0.01$&$0.39\pm0.01$ & $1.81\pm0.04$&Current material\\
PI(75 $\muup$m)+epoxy(5 $\muup$m)& $<2.86 \times 10^{-2}$&$<2.98\times10^{-3}$ & $<6.77\times10^{-3}$&New material\\
\hline
\end{tabular}
}
\label{tab:LA_measurment_result}
\end{table}

The largest risk involved in changing the materials is to maintain the alignment of layers with different thermal expansion coefficients.
In 2015, we successfully fabricated such a low-$\alphaup$ $\muup$-PIC having a detection area 10 $\times$ 10 \si{\cm^{2}}.
The anode electrodes were placed in the centers of the cathode electrodes.
The electrode alignment did not deteriorate when we replaced the insulating material.
We evaluated the anode-voltage dependence and found that the gas gain of the low-$\alphaup$ is comparable to that of a standard $\muup$-PIC.
After evaluating the performance of the fabricated low-$\alphaup$ $\muup$-PIC, we developed a low-$\alphaup$ $\muup$-PIC with a detection area of 30 $\times$ 30 \si{\cm^{2}} in 2016.
The increased size produced no disadvantages.
The device alignment was controlled with a precision of 1 $\muup$m or less.
We checked the gas multiplication of the 30 $\times$ 30 \,\si{\cm^{2}} low-$\alphaup$ $\muup$-PIC and plan to check its anode-voltage dependence.

\section{SURFACE $\alphaup$-RAY DETECTION USING $\muup$-TPC}
\begin{figure}[ht]
	 \centerline{\includegraphics[width=90mm,bb=0 10 725 541]{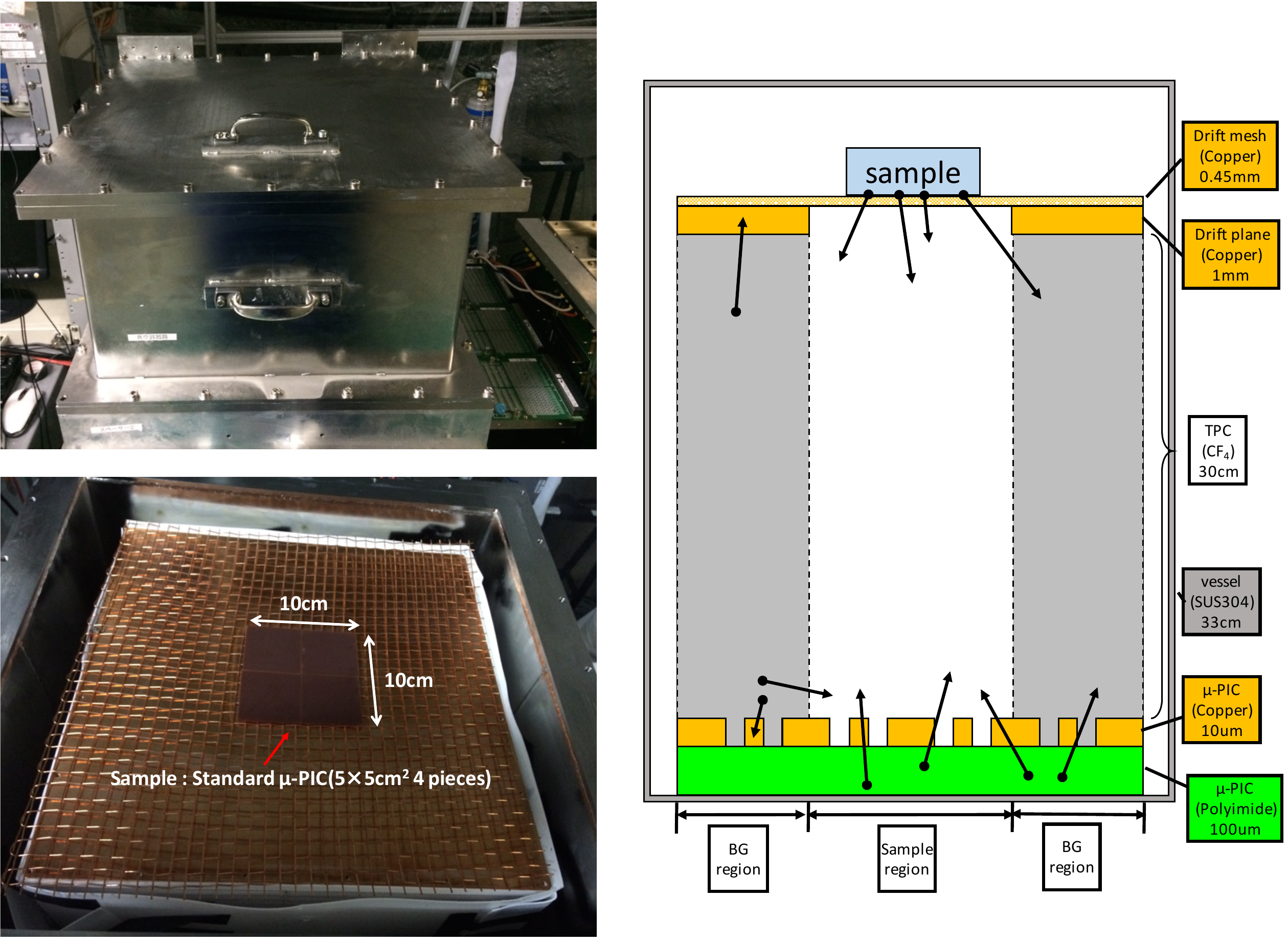}}
  \caption{Surface $\alphaup$-ray detector using a $\muup$-TPC. Upper left : Overview of the surface $\alphaup$-ray detector. Lower left : 10 $\times$ 10 \,\si{\cm^{2}} sample region. Right : Schematic of the surface $\alphaup$-ray detector.}
  \label{fig:Surface_alpha_detector_gaikan}
\end{figure}

The measurement of surface $\alpha$-rays is a significant issue in the search for rare events.
A commercially available instrument UltraLo-1800 developed by XIA are widely used for these measurements\cite{UltraLo}.
It is an ion chamber that uses the pulse shape to discriminate between $\alphaup$-rays produced by the sample from those constituting the background.
Measurement of the insulator samples, if not completely impossible, requires special attention because the pulse shapes may differ from the ideal case due to a deteriorated electric field. 
To measure the surface $\alphaup$-rays, we used a $\muup$-TPC read out by a 30 $\times$ 30 \,\si{\cm^{2}} sized standard $\muup$-PIC, with a detection volume of 30 $\times$ 30 $\times$ 30\,\si{\cm^{3}} filled with CF$_4$ gas at 0.2 atm.
FIGURE\ref{fig:Surface_alpha_detector_gaikan} shows an overview image and the schematic structure of the surface $\alphaup$-ray detector in the Kamioka underground laboratory A.
The $\muup$-TPC is placed in a stainless steel vessel.
The drift plane contains a 10 $\times$ 10 \,\si{\cm^{2}} hole as the sample region.
Samples to be measured are placed on the drift mesh.
A negative voltage is supplied to the drift mesh and the plane.
The sample region is shown in FIGURE\ref{fig:Surface_alpha_detector_gaikan} (lower left).
Our detector can measure insulator samples because the TPC is relatively robust against the deterioration of the electric field.

\begin{figure}[ht]
	 \centerline{\includegraphics[width=110mm,bb=0 0 750 384]{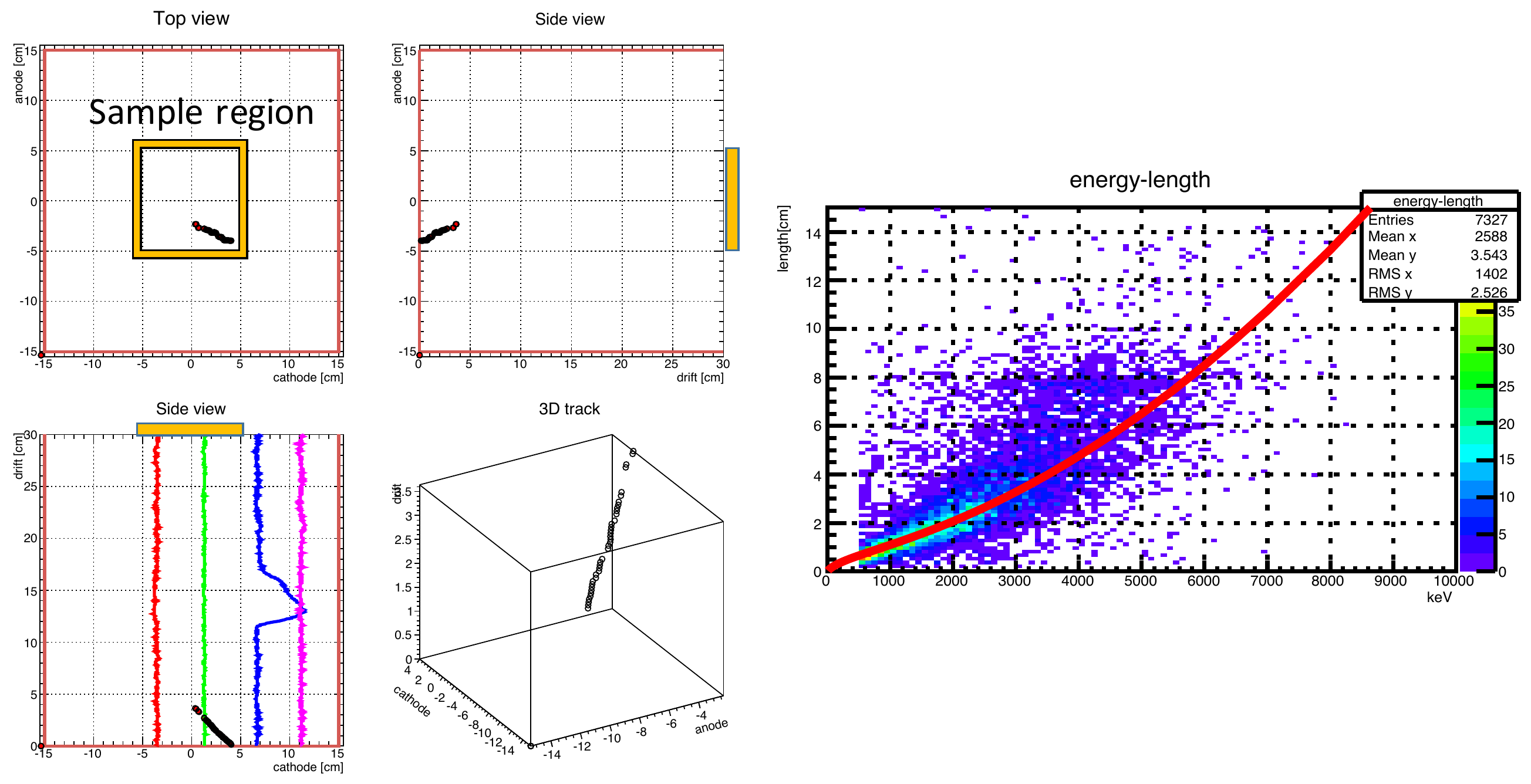}}
  \caption{Left: An example of the event display. The black-circle points represent the hit information. The red, green, blue, and magenta lines are the recorded waveforms. The red line is a fiducial line. Right: Energy dependence of the track length. The red line is a calculated line for $\alphaup$-rays obtained using by SRIM\cite{SRIM}.}
   \label{fig:event_display}
\end{figure}

The $\muup$-PIC has 768 anode strips and 768 cathode strips with a pitch of 400 $\muup$m.
Track and the charge information are recorded for each event. 
A track comprises successive ``hits" in coincidence on both digitized anode and cathode.
For charge information, we grouped the analog signals from the 768 cathode strips into four channels and recorded their waveforms on a flashADC with a 100-MHz clock.
Energy deposition by a charged particle is determined from the waveform.
An example of the event display is shown in the left-hand panel of FIGURE \ref{fig:event_display}.
The yellow and red boxes outline the sample and fiducial regions, respectively.
The black-circle points represent hit information.
The red, green, blue, and magenta lines are the recorded waveforms.

%
%
\subsection{Sample measurement}
\begin{figure}[ht]
	 \centerline{\includegraphics[width= 110mm,bb=0 10 741 553]{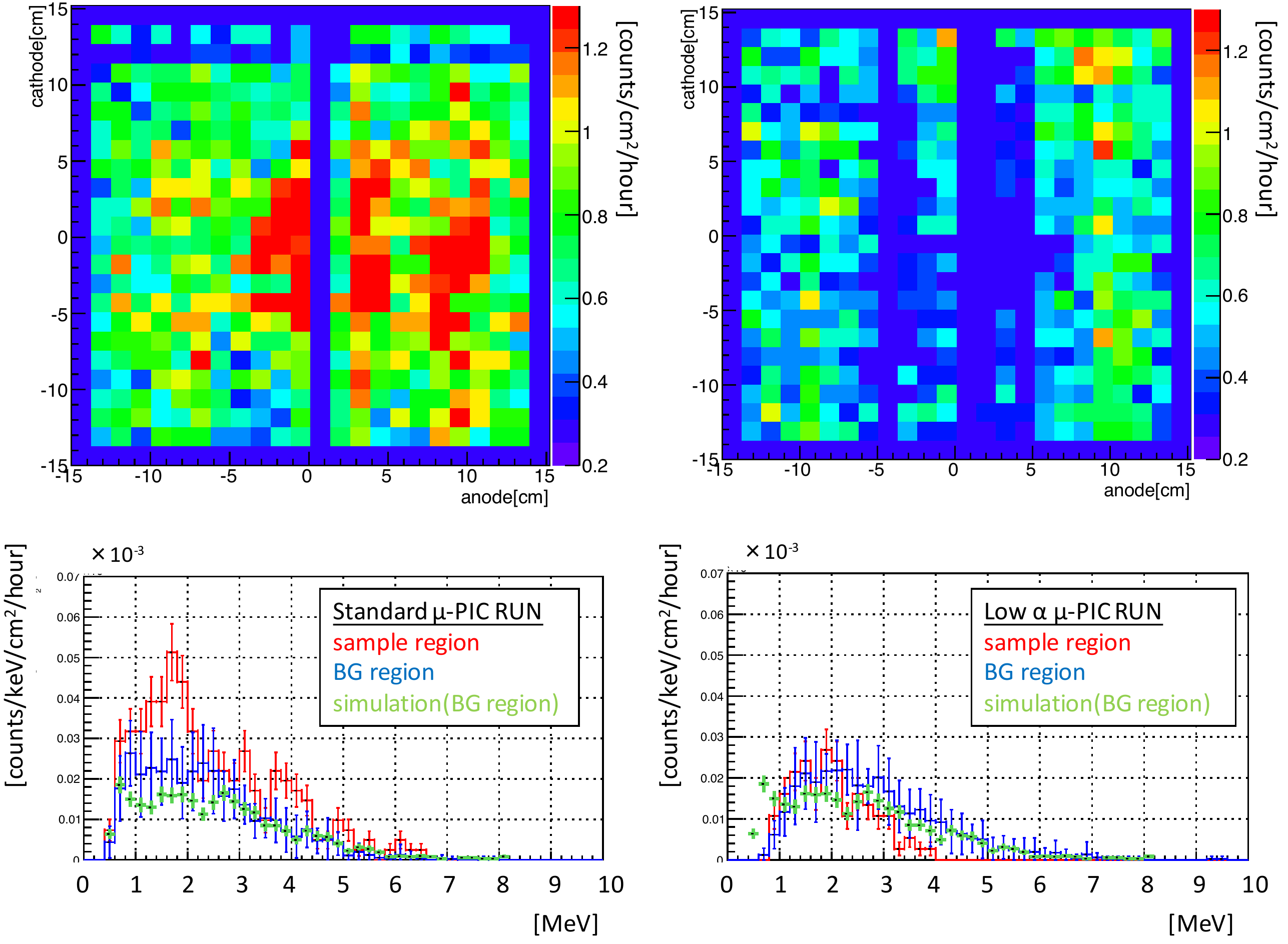}}
  \caption{Upper left : Accumulated image of $\alphaup$-ray tracks for a standard $\muup$-PIC sample measurement. Upper right: Low-$\alphaup$ $\muup$-PIC sample measurement. Lower: The red and blue histograms are the energy spectra for the sample and BG regions, respectively. The green histogram is a Geant4 simulation of $\alphaup$-rays from $^{238}$U and $^{232}$Th in the glass cloth inside the PI100 $\muup$m insulator.}
  \label{fig:2Dhisto_track_start_end}
\end{figure}

We measured the surface $\alphaup$-rays from the $\muup$-PIC samples with our surface $\alphaup$-ray detector to demonstrate the detector performance and measure the $\alphaup$-ray emission of the $\muup$-PIC samples. 
The measured exposure times for four pieces of $5 \times 5\,\si{\cm^{2}}$ samples from a standard and low-$\alphaup$ $\muup$-PIC were 3.16 and 2.88 days, respectively.
We selected events with energies greater more than 500\,\si{keV} and over three hits.
The correlation of energy vs track length from a standard $\muup$-PIC measurement is shown in the right-hand panel of FIGURE \ref{fig:event_display}.
The red line is calculation for $\alphaup$-rays from SRIM\cite{SRIM}, showing that $\alphaup$-rays are clearly detected.
The starting and ending points of the tracks for each event are shown as accumulated images in the upper panels of FIGURE \ref{fig:2Dhisto_track_start_end}.
The lower panels of FIGURE \ref{fig:2Dhisto_track_start_end} show the energy spectra for the sample and background regions.
The image loss observed in the upper left-hand along the anode between 0 and 1 cm, which we denote by $0\,\si{\cm} <$ anode $< 1\,\si{\cm}$, and between $-5\,\si{\cm}<$ anode $<-4\,\si{\cm}$ and $0\,\si{\cm}<$ anode $<5\,\si{\cm}$ in the upper right-hand panel are due to the faulty connections between the detector and electronics.
These regions were removed from the surface $\alphaup$-ray analysis.
The area between $3.5\,\si{\cm}<$ anode $<-0.5\,\si{\cm}$ and $-4.5\,\si{\cm}<$ cathode $<4.5\,\si{\cm}$( 3 $\times$ 9 \,\si{\cm^{2}} area designate as ``1 block") is defined as the sample region, and the rest is defined as the background (BG) region.
The BG region is segmented into 14 blocks.
The count rate for a standard $\muup$-PIC sample is 0.113$\pm$0.007 [counts /$\si{\cm^{2}}$/h], where the error is a statistical error.
The corresponding count rate for the BG region is 0.068$\pm$0.007$\pm$0.020 [counts/$\si{\cm^{2}}$/h], where the first and second error terms are a statistical and systematic errors, respectively.
The position-dependence of the BG region blocks is a systematic error.
The energy spectrum for each region is shown in FIGURE \ref{fig:2Dhisto_track_start_end} (lower panels).

We obtained the count rates for the $\alphaup$-rays from the samples by subtracting the rate for the BG region rate from that for the sample region.
The rate for the surface $\alphaup$-rays from the standard $\muup$-PIC sample is 0.034$\pm$0.009 [counts/$\si{\cm^{2}}$/h].
We evaluated the detection efficiency for surface $\alphaup$-rays as 0.159$\pm$0.007 using Geant4, where the error is the systematic error due to the uncertainly in the glass-cloth thickness.
The rate of surface $\alphaup$-rays from a standard $\muup$-PIC sample is 0.28$\pm$0.12 [$\alpha$ /$\si{\cm^{2}}$/h].
For comparison, the value determined from the HPGe measurements is 0.146$\pm$0.004 [$\alpha$ /$\si{\cm^{2}}$/h], where the error includes the systematic error due to the uncertainly in the glass-cloth thickness and the statistical error.
These values are consistent at about the 1.1$\sigmaup$ level.
The rate of surface $\alphaup$-rays for the low-$\alphaup$ $\muup$-PIC sample was analyzed in the same way as for the standard $\muup$-PIC sample.
No significant excess over the background was detected, setting a 90$\%$ upper limit of $7.55 \times 10^{-2}$  [$\alpha$ /$\si{\cm^{2}}$/h].

\begin{table}[ht]
\centering
\caption{Surface $\alphaup$-ray measurements for different samples compared with HPGe measurements. For surface $\alphaup$-rays, we consider the statistical error. For HPGe measurements, the error include the statistical error and systematic errors due to the glass-cloth thickness uncertainly.}
\begin{tabular}{l|c|c}
\hline
Sample & Measurement result[$\alphaup$/$\si{\cm^{2}}$/h] & Estimated value from HPGe measurement [$\alphaup$/$\si{\cm^{2}}$/h]\\ \hline 
Standard $\muup$-PIC&$0.28\pm0.12$& 0.146 $\pm$ 0.004 \\
Low $\alphaup$ $\muup$-PIC & $<7.55 \times10^{-2}$&$<10^{-4}$\\
\hline
\end{tabular}
\label{tab:surface_alpha_measurment_result}
\end{table}

To understand the background for the surface $\alphaup$-ray detector, we simulated the energy contributions owing to $\alphaup$-rays from $^{238}$U and $^{232}$Th in the glass cloth inside the PI100 $\muup$m insulator of the $\muup$-PIC using Geant4 (FIGURE \ref{fig:2Dhisto_track_start_end}).
The background is well-explained by the $\alphaup$-rays from $^{238}$U and $^{232}$Th in the glass cloth inside the PI100 $\muup$m insulator. 
We also determined the background contribution from radon contamination in the gas. 
Assuming that events with energies of at least 4$\,\rm{MeV}$ are the $\alphaup$-rays from radon, we obtained $ 0.66\,[\si{\becquerel/m^3}]$ as the upper limit for radon events.
Considering the detection efficiency for the radon events, we found the upper limit to the radon rate as $1.1[\alpha/ \si{\cm^{2}}$/h].
\par
We plan to replace the standard $\muup$-PIC as a read-out device with a low-$\alphaup$ $\muup$-PIC and install a gas-circulation system with the a cooled charcoal filter.
After replacing the $\muup$-PIC, we expect the BG from the $\muup$-PIC to be reduced to less than $10^{-4}$ [$\alphaup$/$\si{\cm^{2}}$/h].
Previous studies show that the radon rate can be reduced by a factor $\sim 1/50$ using the gas-circulation system, decreasing the BG from radon to less than $\sim0.02$ $[\alphaup/ \si{\cm^{2}}$/h].
\section{CONCLUSION}
NEWAGE is a direction-sensitive, dark-matter-search experiment using a gaseous micro-time projection chamber.
We achieved a 90\% CL direction-sensitive SD cross-section limit of 557\,\si{pb} for a WIMP mass of 200\,\si{GeV/c^{2}}.
We found the dominant background to be $\alphaup$-rays from the glass cloth used to reinforce the PI100 $\muup$m insulator in the $\muup$-PIC.
Accordingly, we developed a low-$\alphaup$ $\muup$-PIC by replacing the top layer of PI with a new material we developed.
We plan to check the anode-voltage dependence of the gas gain of the 30 $\times$ 30 \,\si{\cm^{2}} low-$\alphaup$ $\muup$-PIC.
We used a $\muup$-TPC for surface $\alphaup$-ray measurements of both standard and low-$\alphaup$ $\muup$-PIC samples.
The value of the surface $\alphaup$-rays from the standard $\muup$-PIC is consistent within about 1.1$\sigmaup$ of the value determined from the HPGe measurement.
We found the dominant background for the surface $\alphaup$-ray detector to be $\alphaup$-rays from the $\muup$-PIC used as the read-out device.
If the $\muup$-PIC is replaced, the background level will be limited by the radon in gas.
We plan to replace the standard $\muup$-PIC as the read-out device by a low-$\alphaup$ $\muup$-PIC and introduce a gas-circulation system with a cooled charcoal filter.
\section{ACKNOWLEDGMENT}
This study was partially supported by KAKENHI Grant-in-Aids(26104001,26104005,26104008 and16H02189).
This work was partially carried out by the joint research program of the Institute for Cosmic Ray Research (ICRR), University of Tokyo.
We gratefully acknowledge the ICRR staff, the cooperation of the Kamioka Mining and Smelting Company.



\begin{thebibliography}{99}
\bibitem{Planck}Planck Collaboration, A\&A 571 (2014) A16.
\bibitem{DAMA_2012}R. Bernabei et al., J. Phys. Conf. Ser. 375 (2012) 012002.
\bibitem{XENON}E. Aprile et al. (XENON100 Collaboration), Phys. Rev. Lett. 109, 181301.
\bibitem{LUX}D.S. Akerib et al. (LUX Collaboration), Phys. Rev. Lett. 118, 021303.
\bibitem{PANDA}Y. Yang, On behalf of PandaX-II Collaboration, arXiv:1612.01223v1.
\bibitem{DM_direction_strong_evidence}D.N. Spergel, Phys. Rev. D 37 (1988) 1353.
\bibitem{NIT_NIM2007}T. Naka et al., Nucl. Instrm. Methods Phys. Res. Sect. A 581 (2007) 761.
\bibitem{CNT_DM}L.M. Capparelli et al., Physics of the Dark Universe 9-10 (2015) 24-30.
\bibitem{DRIFT_APP2012}E. Daw et al., Astropart. Phys. 35 (2012) 397.
\bibitem{DMTPC_PLB2011}S. Ahlen et al., Phys. Lett. B 695 (2011) 124.
\bibitem{MIMAC_2011}D. Santos et al., J. Phys. Conf. Ser. 309 (2011) 012014.
\bibitem{uPIC}A. Takada et al., Nucl. Instr. Meth. Phys. Res. Sect. A 573, (2007) 195.
\bibitem{GEM}F. Sauli and A. Sharma, Annu. Rev. Nucl. Part. Sci. 49, (1999) 341.
\bibitem{NEWAGE_2015}K. Nakamura et al., Prog. Theor. Exp. Phys. (2015) 043F01.
\bibitem{Geant4}S. Agostinelli et al., Nucl. Instr. Meth. Phys. Res. A 506 (2003) 250.
\bibitem{UltraLo}W. K. Warburton, J. Wahl, and M. Momayezi, Ultra-low Background Gas-filled Alpha Counter, U.S. Patent 6 732 059, May 4, 2004. W. K. Warburton et al., IEEE Nuclear Sci. Symp. Conf. Rec., Oct. 16-22, 2004, vol.1, pp. 577-581, Paper N16-80. M. Z. Nakib et al., AIP Proc. 1549, 78-81
\bibitem{SRIM}J.F. Ziegler, J.P. Biersack SRIM The Stopping and Range of Ions in Matter, Code (1985).
\end{thebibliography}
\end{document}